# Suppression of Vertical Oscillation and Observation of Flux Improvement during Top-up Injection at PLS-II

Y-G. Son,[1] J.-Y. Kim,[1] C. Mitsuda,[2] K. Kobayashi,[2] J. Ko,[1] T-Y. Lee,[1] J-Y. Choi,[1] D-E. Kim,[1] H-S. Seo,[1] H-S. Han,[1] K-S. Park,[1] and S. Shin[1,3]

[1] Pohang Accelerator Laboratory, POSTECH, Pohang, Kyungbuk, 790-784, KOREA
[2] SPring-8, Kouto, Sayo-cho, Sayo-gun, Hyogo, 679-5198, JAPAN
[3] Advanced Photon Source, Argonne National Laboratory, Illinois 60439 USA

E-mail: tlssh@postech.ac.kr

**Abstract**. This paper reports a start-to-end study of suppression of stored beam oscillation at PLS-II. We report that the fast counter-kicker implemented in PLS-II suppressed vertical oscillation of the stored beam. During top-up injection in the magnetic spectroscopy beamline of PLS-II, the stored beam oscillation was suppressed by a factor of nine, and flux was improved by a factor of three.

## 1. Introduction

By successes at APS [1] and SLS [2], top-up operation is becoming the standard mode of operation in most of the third generation light source. Therefore, constant beam current in the storage ring to overcome lifetime limitations and a constant photon flux for a thermal equilibrium at the beamline have been realized by the help of top-up operation. PLS-II [3] has been successfully performing top-up operation since March 2013. As a result, the electron beam current has been regulated with rms 0.09 % at 400 mA, and the flux change for 8 h has improved from 40 % in decay mode to 2 % in top-up operation at the supramolecular crystallography beam line [4].

During the top-up operation, the beam injection system excites an oscillation of the stored beam. The excited oscillation effectively enlarges the stored beam emittance and modulates the photon beam intensity. For example, 1~8 % systematic dips in the photon beam flux pattern of each beam line occur during the top-up injection in PLS-II. Furthermore, future storage ring-based light sources will have ultra-low emittance. In these light sources, the lifetime of a stored electron beam will be extremely short and the frequency of beam injection for top-up will increase. Therefore, to achieve high performance top-up operation for user experiments in the future low emittance ring, the stored beam oscillation must be suppressed during top-up injection.

In general, there is common origin for the transverse oscillation of the stored beam during injection. That is imperfection of main kicker system such as tilt of bump magnets, the non-similarity of the magnetic fields of the bump magnets and waveform errors for the pulsed kicker magnets. Other origins of the stored beam oscillation depend on inherent injection system; e.g., nonlinearity within an injection bump orbit, and leakage field from septum magnets. One effort to suppress stored beam oscillation has been concentrated on improving the imperfection of each injection system. The other effort is to implement compensator for the imperfection of injection system. Beside these efforts to suppress stored beam oscillation, a new injection scheme using a single pulsed quadrupole magnet without local bump was proposed and demonstrated at the PF-KEK [5].

The injection region in PLS-II includes four kickers for an orbit bump; they are located within the

6.8 m long straight section. Therefore, the sextupole magnets do not cause injection bump leakage. The kicker magnet is operated at 10-Hz repetition rate and has a 7 μs half-sine wave so that the rise and the fall times are both 3.5 μs. The incoming electron beam from the beam transfer line is horizontally parallel to the bumped orbit in the storage ring, and is injected 8.234° vertically. A Lambertson magnet then bends the beam -8.234° vertically to place the incoming electron beam on the same plane as the bumped orbit. The main sources of error on the horizontal and vertical stored beam oscillations are waveform errors for pulsed kicker magnets and the leakage field of the Lambertson septum magnet, respectively. As a result, 1~8 % systematic dips in the photon beam flux pattern of each beam line occur during beam injection in PLS-II.

Figure 1 (a) and (c) shows a model of the Lambertson septum magnet [6] for PLS-II and the distribution of the leakage $B_x$ along the z-axis from a field measurement, respectively. The measured distribution of the leakage $B_x$ along the z-axis is less than the 1000 G·cm tolerance specification, and the difference between the field at the stored beam's position and the bumped orbit's position is 230 G·cm in Fig. 1(c). Particularly, the measured leakage $B_x$ field integral decreased changes linearly along the x-axis, possibly as a result of an air gap in the V-notch area. Various restrictions limit the amount by which the leakage field can be reduced. Therefore, rather than removing the oscillation source, a fast counter-kicker is implemented to compensate for stored beam oscillation by the Lambertson septum magnet in PLS-II. As a result, in the magnetic spectroscopy beam line of PLS-II, the vertical oscillation of the stored beam is suppressed by a factor of nine, and flux is improved by a factor of three [7].

In this paper we describe in detail the stored beam oscillation due to the imperfection of the injection system, report the result of the suppression of stored beam vertical oscillation by the fast counter kicker implemented in PLS-II, and observe the improvement of flux in PLS-II beam line during top-up injection. Section 2 introduces the theoretical description on stored beam oscillation during top-up injection. Section 3 describes the development of a fast counter kicker. Section 4 describes the measurement of stored beam oscillation, the experimental result of the oscillation suppression, and flux improvement by the fast counter-kicker. Section 5 presents conclusions.

## 2. Beam oscillation during injection

To describe the perturbation caused in the stored beam by the imperfection of the injection system, we can use a one-turn transfer matrix $T$ to track the particle motion [8]:

$$\begin{pmatrix} x \\ x' \\ y \\ y' \end{pmatrix}_n = T \begin{pmatrix} x \\ x' + \theta_x \\ y \\ y' + \theta_y \end{pmatrix}_{n-1}, \quad (1)$$

where $n$ is the index for the turn number, and $\theta$ represents the kick induced by the imperfection of the injection system. Here the full-turn coupled transfer matrix $T$ is decomposed into normal modes as follows

$$T = \begin{pmatrix} M & n \\ m & N \end{pmatrix} = VUV^{-1}. \quad (2)$$

where

$$U = \begin{pmatrix} A & 0 \\ 0 & B \end{pmatrix}; \quad V = \begin{pmatrix} \gamma I & C \\ -C^\dagger & \gamma I \end{pmatrix} \text{ and } \gamma^2 + |C| = 1. \quad (3)$$

$$A = \begin{pmatrix} \cos 2\pi v_A + \alpha_A \sin 2\pi v_A & \beta_A \sin 2\pi v_A \\ -\gamma_A \sin 2\pi v_A & \cos 2\pi v_A - \alpha_A \sin 2\pi v_A \end{pmatrix} \quad (4)$$

is the full-turn transfer matrix for one of the normal modes; $B$ is written similarly. $T$, $U$ and $V$ are 4 x 4 matrices. $A$, $B$ and $C$ are 2 x 2 matrices. $I$ is the 2 x 2 identity matrix.

In PLS-II, $\theta_x$ is mainly caused by non-similarity of field shape in each kicker magnet and $\theta_y$ is mainly caused by the leakage field in the septum magnet. The vertical kick strength caused by the leakage field can be converted by the integrated leakage field in Fig. 1 (c) as $\theta_y \ [rad] = 0.29979 \times Bl \ [T \cdot m] / E \ [GeV]$ [9]. While the injection bump is being constructed, the stored beam's position is changed along the x-axis in Fig. 1 (c). Linearly fitting the data in the figure and using the relation between the field integral and the kick, yields the relation between the kick strength and the x-axis:

$$\theta_y[rad] = -0.00147 \times h[m], \tag{5}$$

where $h$ is the bump's height in the injection region. Because kickers $K_1$, $K_2$, $K_3$, and $K_4$ are symmetrically located on both sides of the injection point and no optical elements (e.g., quadrupole magnet, sextupole magnet) are placed among kickers so that a bump orbit can be formed in PLS-II, the bump height can be simplified as [10]

$$h[m] = -\theta_k(n)[rad] \times L_{12}[m], \tag{6}$$

where $n$ is the index for the turn number, $\theta_k$ is the deflection angle by the kicker, and $L_{12}$ = 1590 mm is center-to-center distance between $K_1$ and $K_2$. The bump height is proportional to the kicker magnetic field in Eq. (6). The kicker magnet is operated at a 10-Hz repetition rate and has a 7-μs half-sine wave; i.e., the rise and the fall time are both 3.5 μs. The kick to construct the injection bump can be formulated as

$$\theta_k(n,M)[rad] = \theta_0[rad]\sin(\pi \times \frac{t}{T_k}), \tag{7}$$

where $\theta_0$ = 10 mrad is the maximum deflection angle of the kicker, $T_r$ = 1 μs is the revolution period, $T_k$ = 7 μs is the pulse length of the kicker, $1 \leq M \leq 470$ is bunch index, and $0 \leq n \leq 6$ is turn number while the kicker is on. Substituting Eqs. (7) and (6) into Eq. (5) yields the formula to represent the kick by the leakage field while the kicker is on:

$$\theta_k(n,M)[rad] = 0.00147 \times \theta_0[rad]L_{12}[m]\sin(\pi \times \frac{t}{T_k}). \tag{8}$$

Eq. (8) yields the kick caused by the leakage field from the septum magnet. By solving Eq. (1) for every turn, we can estimate the stored beam oscillation during injection. Note that $\theta_x$ can also be considered straightforwardly according to error sources and $\theta_x = \theta_y = 0$ after kicker off (after 7 μs) in Eq. (1).

The particle tracking code ELEGANT [11] was used to perform a detailed study to estimate the effect of leakage field quantitatively and to investigate some techniques and operation schemes to reduce the oscillation's amplitude. To study the effect of leakage field from septum magnet, determined kick in Fig. 2 from Eq. (8) is considered in the ELEGANT simulation. In Fig. 3(a), the stored beam oscillation was represented in phase space at the source point of the magnetic spectroscopy beamline. The stored beam oscillation due to septum leakage field (while the kicker is on and after it turns off) has 100 μm amplitude.

The septum's integrated leakage field has good linearity along the x-axis in Fig. 1(c). Therefore, the vertical oscillation caused by the leakage field may be easily suppressed by imposing a counteracting linear compensation source. By tilting the injection kicker magnets deliberately and systematically [6], the vertical kick from the injection kicker magnets can be used to compensate for the effect of the septum's leakage field. When the tilt angles of the injection kicker magnets are the same and compensate for each other in the horizontal plane, horizontal orbit oscillation does not occur; instead a vertical kick occurs according to the tilt angles. In the simulation, this vertical oscillation could be reduced to 1 μm amplitude by tilting the injection kicker magnets appropriately in Fig. 3 (blue triangles). Adjustment of vertical tune can be also considered to reduce the vertical oscillation caused by the leakage field from the septum magnet. Eq. (1) indicates that while the kicker pulse is on, leakage field kicks along the height of the bump orbit can be additive or subtractive depending on the transfer matrix elements. By changing the vertical tune, we can manipulate each transfer matrix element. In the simulation, the vertical oscillation's amplitude can be suppressed to as low as 20 μm (Fig. 3a, red circle).

The same pulse shape counter kicker [12] with the kick shape by the leakage field from the septum magnet is a good candidate to suppress the stored beam oscillation caused by the leakage field. To simulate the effect of the counter kicker, a counter kicker with the same pulse shape is considered and its amplitude and timing were tuned in ELEGANT. The septum magnet is located in cell 1; as result, the stored beam oscillation is suppressed to < 1 μm in Fig. 3 (green triangle). Because this method is simple (just two knobs for operation) and does not require any changes to beam parameters, use of a counter kicker to suppress beam oscillation by the leakage field is strongly recommended.

### 3. Fast counter kicker system

A fast counter kicker system is needed to suppress stored beam oscillation due to the leakage field from the septum magnet. The required kick angle of the counter-kicker is > 30 μrad with a pulse width 7 μs. To generate the required kick, an air-core magnet and LC resonant power supply have been developed. The fast counter-kicker system in Fig. 4(a) consists of a dipole air-core magnet, a compact driving power supply circuit and an external high-voltage power supply. The ceramic chamber is wound with coils to generate both horizontal and vertical magnetic fields. To keep the impedance low in the stored beam, the inner surface of the ceramic chamber is coated with Ti-Mo of 6 um thickness.

The fast counter kicker is a series resonant circuit (Fig. 4a, Table 1). Output current waveform is a half-sinusoid. Three-layer pulse forming networks (PFNs) are used to form the half-sine waveform, and a 25-m coaxial cable from power supply to chamber magnet is expressed as an equivalent circuit. The power supply of line type consists of many inductors and capacitors and is composed of series and parallel circuit combinations. An event timing system provides trigger signal and timing delay. To realize fast rising time, we used a MOSFET semiconductor switch in Fig. 4(b) to feed a current signal of a sine half wave in the respective coils, and used series resonance in the energy storage capacitor and magnet coil. In these ways it differs from the fast count kicker system in Spring-8 [12]. As shown in Fig. 5, measured output current had pulse duration 7 μs, which is the time required for the suppression; the peak current was 10 A (can be maximized to 60 A).

### 4. Experimental result

To explore the phenomenon of beam oscillation during injection, all 96 Libera Brilliance BPMs [13] were used in the measurement. Two thousand samples from 96 turn by turn BPMs were taken after injection trigger on and analyzed using a singular-value decomposition (SVD) method [14]. In a singular value from a diagonal matrix, two large modes are separable from the floors in the vertical plane, and three large modes are separable from the floors in the horizontal plane. In the vertical plane, the main oscillatory motion is betatron oscillation with 50 μm amplitude; the phases of the betatron oscillatory motion of the two dominant modes differ from each other 90°, as in normal SVD analysis results. In the horizontal plane, energy oscillatory motion occurs in addition to the two betatron motions with 90° phase difference. A 10 kHz energy oscillatory motion occurs due to synchrotron phase offset caused by large orbit deviation while the kicker is on.

The first and second vectors for spatial and temporal modes from SVD analysis show a clear dispersion pattern with horizontal betatron pattern in Fig. 6(a). The first temporal mode shows large horizontal orbit deviation while the kicker is on, and the second temporal mode shows 10 kHz energy oscillatory motion in Fig. 6(b). The measurements show large deviation (2 mm) and short time duration (7 μs), so the source of horizontal motion should be suppressed rather than compensated for. Remarkably, horizontal large deviation while the kicker is on can be also used to compensate for vertical oscillation by manipulating the coupling term in Eq. (1) [15].

To suppress vertical oscillatory motion, the fast counter kicker system has been implemented in cell 7 of PLS-II storage ring. First, the polarity and trigger timing for beam oscillation by fast counter kicker system was calibrated. Then the improved oscillation phase and amplitude of the fast counter kicker were scanned to suppress vertical oscillatory motion minimally. This procedure to find the optimal setting is simple and straight forward because it uses only two knobs for tuning. As result, the stored beam oscillation was suppressed by the factor of nine. Phase space beam motion at the 2A Beamline (magnetic spectroscopy beamline) during injection with and without fast counter kicker show that the effect of the counter kicker is negligible during the first 7 μs (kicker waveform duration) (Fig. 7b) but the counter kicker took 7 μs (after the kicker was turned off) to suppress the oscillation by a factor of nine, because of offset and angle error caused by the kick from the leakage field at the counter kicker position; suppression of these errors required several turns.

The photon flux improvement was also measured at the 2A magnetic spectroscopy beamline in Fig. 8. The photon source of the beamline is an Elliptically Polarized Undulator (EPU) with the magnet period of 7.2 cm. The beam flux improvement was measured with a gold mesh just after the exit silt with a vertical gap of 10 μm. At the exit slit position, the FWHM of the monochromatic central cone is also about 10 μm, making the mesh current very sensitive to variation of beam flux. The data acquisition was at a rate of 600 Hz, which is enough to monitor the 10 Hz injection transient. By help of the fast counter kicker, the dip of photon beam flux due to injection oscillation was improved by a factor of three in Fig. 9.

## 5. Conclusion

By implementing the fast counter kicker system in PLS-II, we suppressed vertical oscillation of stored beam by a factor of nine during top-up injection. As result, photon beam flux was improved by a factor of three in the magnetic spectroscopy beamline of PLS-II. These results demonstrated that the fast counter kicker system is promising candidate to suppress stored beam oscillation caused by systematic error, and to increased beamline flux during top-up injection. However, stored beam oscillation by complex waveform error should be reduced by suppressing the error source rather than by compensating for the error.

**Acknowledgments**
We would like to H. Wiedemann (SLAC) and L. Emery (APS) for useful discussions. This research was supported by the Basic Science Research Program through the National Research Foundation of Korea (NRF-2015R1D1A1A01060049).

**References**
[1] L. Emery, Proc. of PAC01, p. 2599.
[2] J. Bengtsson *et al*., Proc. of EPAC96, p. 685.
[3] S. Shin *et al*., JINST 8, P01019 (2013).
[4] I. Hwang *et al*., Rev. Sci. Instrum. 85, 055113 (2014).
[5] Kentaro Harada *et al*., Phys. Rev. ST Accel. Beam 10, 123501 (2007).
[6] D-E. Kim *et al*., Journal of the Korean Physical Society, p. 197-204, v. 64 (2014).
[7] The magnetic spectroscopy beamline in PLS-II; *http://pal.postech.ac.kr/paleng/bl/2A/*
[8] P. Bagley *et al*., Proc. of PAC89, p. 874 (1989).
[9] H. Wiedemann, "Particle Accelerator Physics", Springer, Berlin, Heidelberg, New York (2007).
[10] Kuanjun Fan *et al*., NIM. A 450 p. 573 (2000).
[11] M. Borland, "elegant: A Flexible SDDS-Compliant Code for Accelerator Simulation," Advanced Photon Source LS-278, September 2000.
[12] C. Mitsuda *et al*., IPAC10, Kyoto, Japan, p. 2552 (2010).
[13] http://www.i-tech.si.
[14] C.-X. Wang, PAC2003, Portland, OR, USA, p 3410 (2003).
[15] I. Hwang, *private communication*.

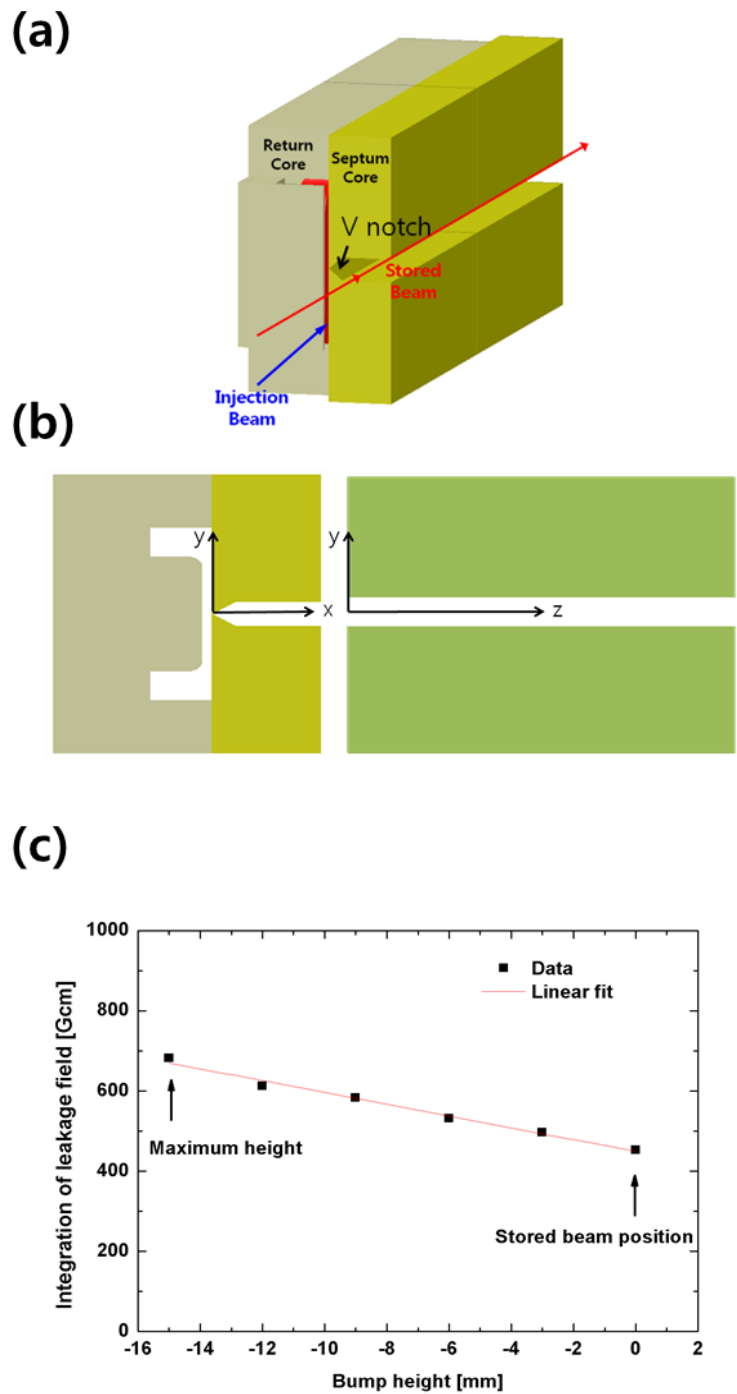

**Figure 1.** (a) Model of the septum magnet. (b) Definition of coordinates along septum magnet. (c) Integrated leakage field along the bump height (x-axis).

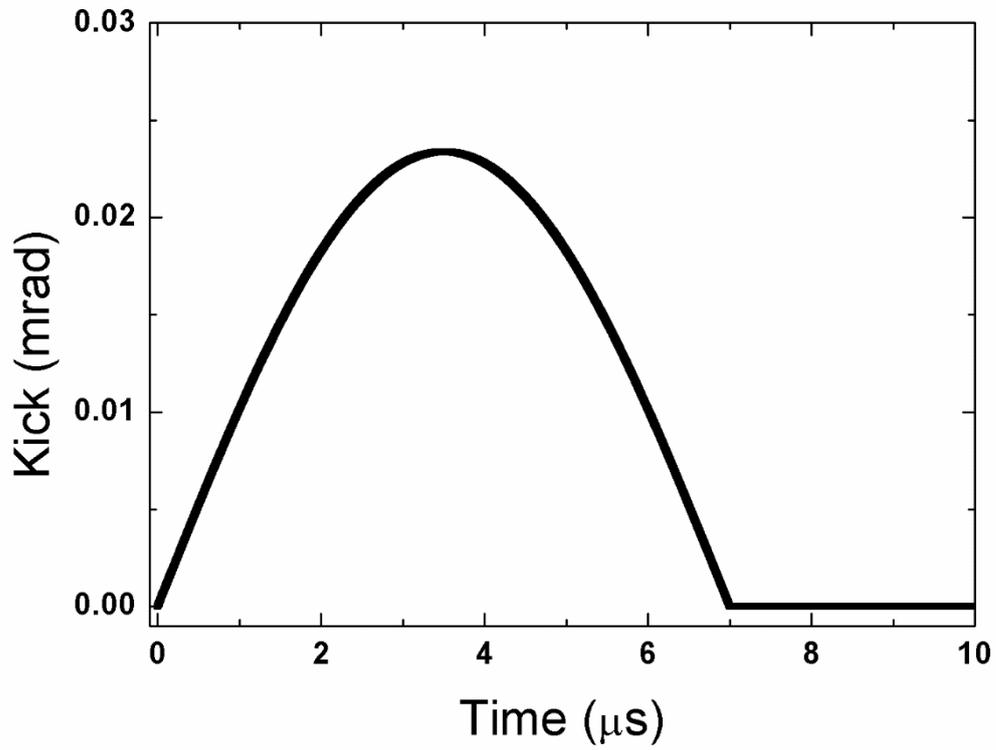

**Figure 2.** Kick caused by septum magnet. This kick is generated due to variation in the septum leakage field along the horizontal injection bump. Therefore, the shape of the kick depends on the field profile of the injection kicker.

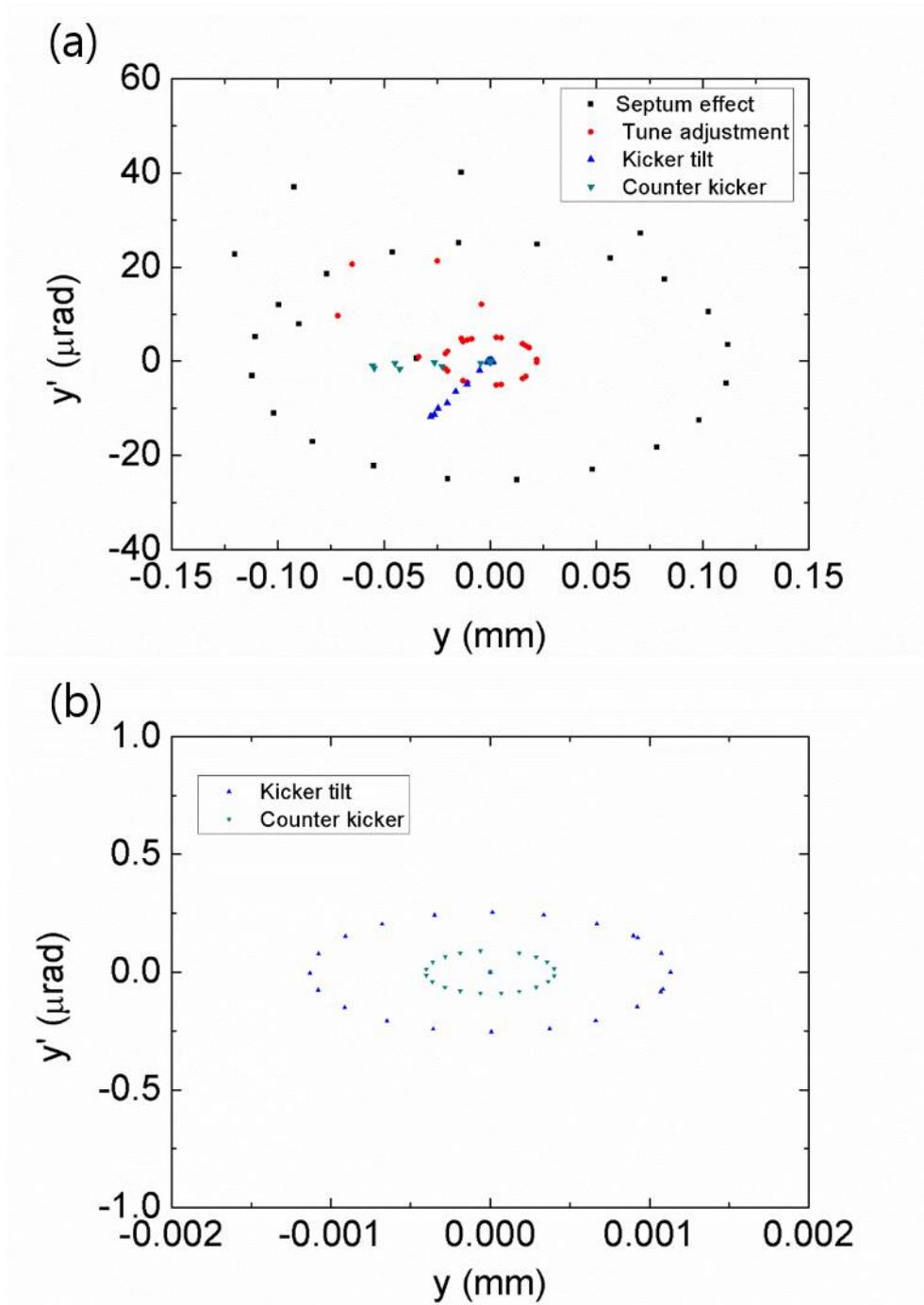

**Figure 3.** Stored beam motion in phase space at the source point of long insertion device (cell 2). (a) The oscillation due to septum leakage field. (b) The suppressed oscillation by counter kicker located at cell 2.

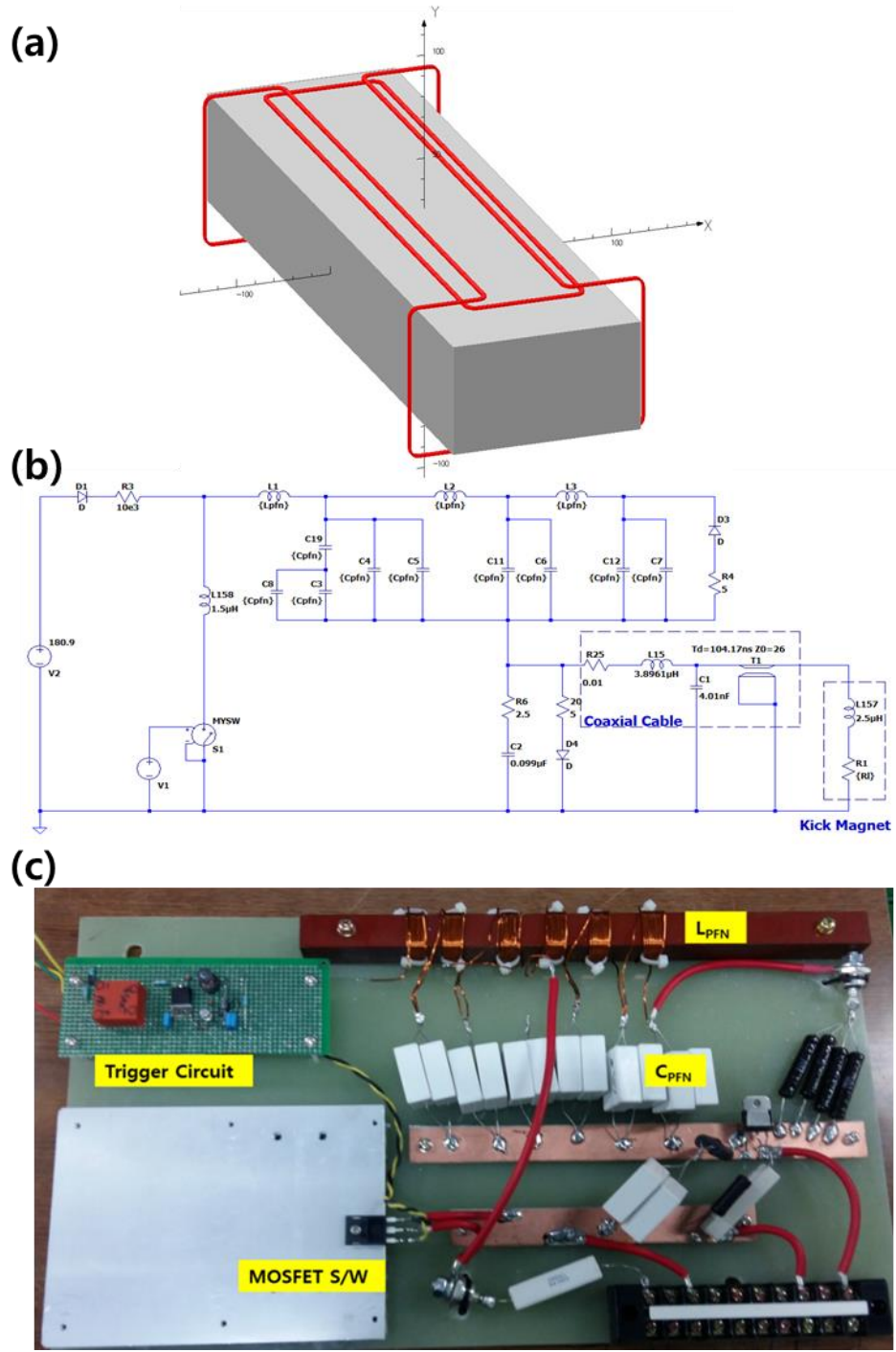

**Figure 4.** Fast counter kicker system. (a) The schematic of ceramic chamber. (b) Circuit diagram for electric circuit simulation. (c) Photograph of power supply for the fast counter kicker.

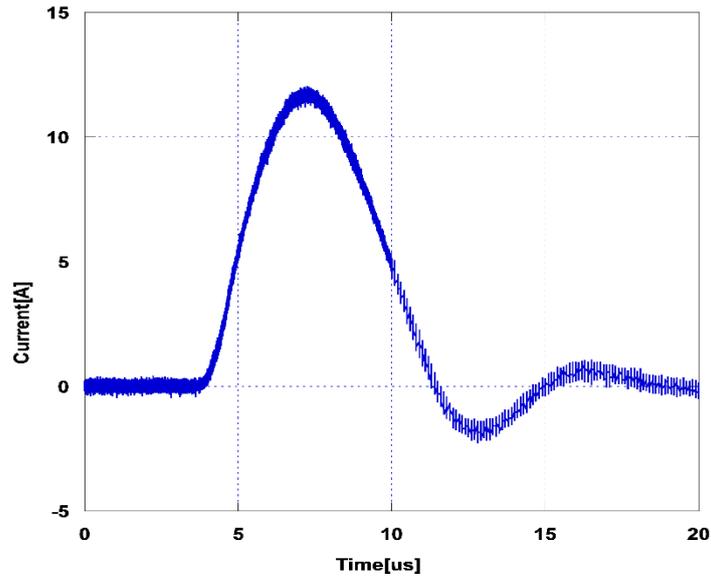

**Figure 5.** Measured output current waveform.

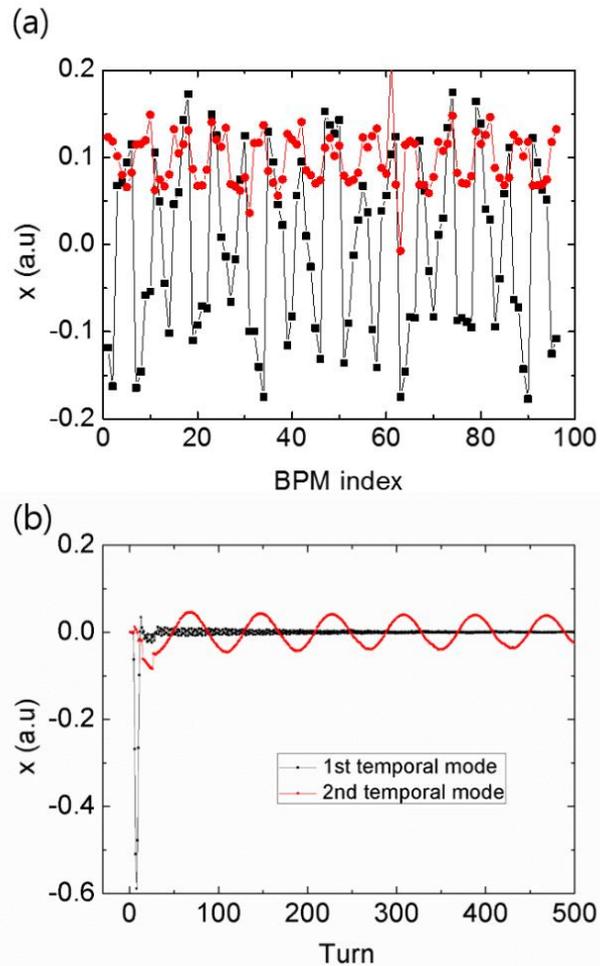

**Figure 6.** (a) 1st and 2nd spatial modes from SVD analysis. 1st mode (black square): horizontal betatron oscillation; 2nd mode (red circle): dispersion motion. (The 61st and 63rd BPMs malfunctioned.) (b) 1st and 2nd temporal modes from SVD analysis. 1st mode (black square): horizontal betatron oscillation with large amplitude while the kicker is on; 2nd mode (red circle): dispersion motion with 10-kHz synchrotron frequency.

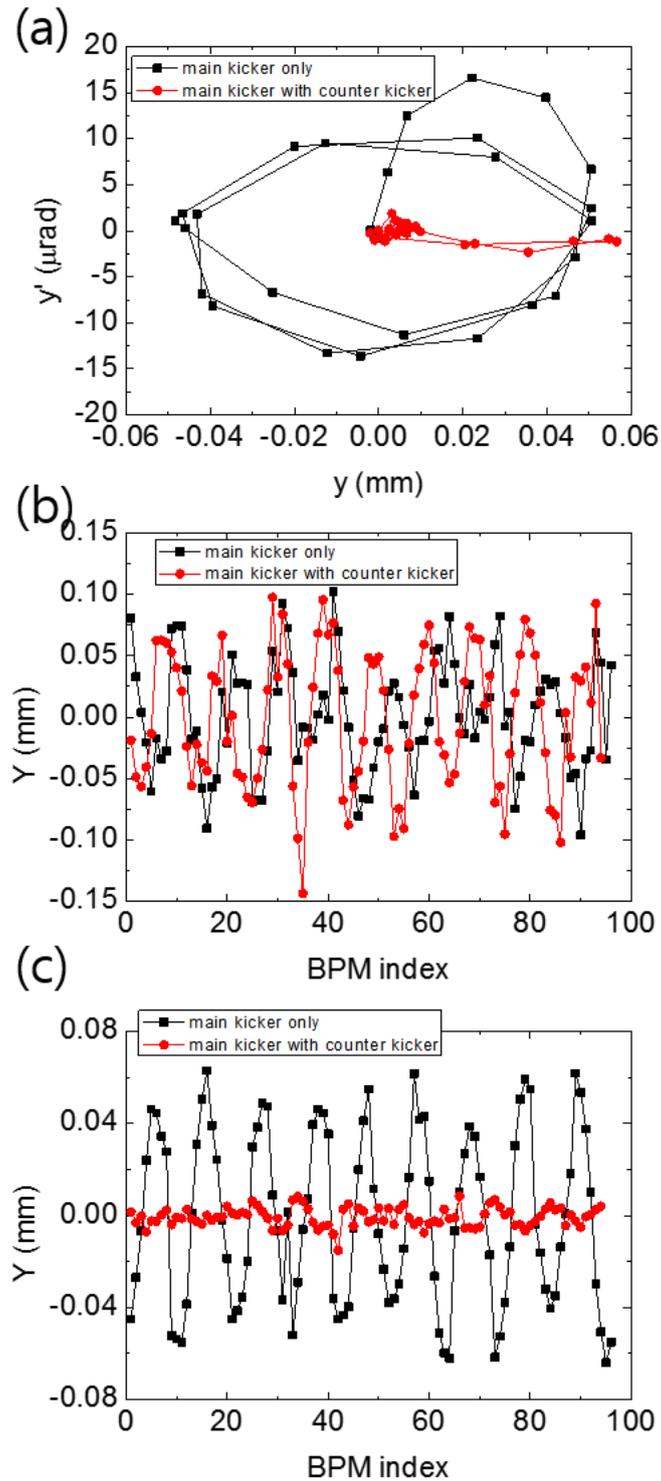

**Figure 7.** (a) Measured phase space beam motion at 2A BL (magnetic spectroscopy beamline) with and without counter-kicker. (b) Measured beam oscillation along all BPMs for 4th turn after the kicker was turned on: oscillation was not suppressed while the kicker was on. (c) Measured beam oscillation along all BPMs while the kicker was of: oscillation was suppressed by a factor of nine.

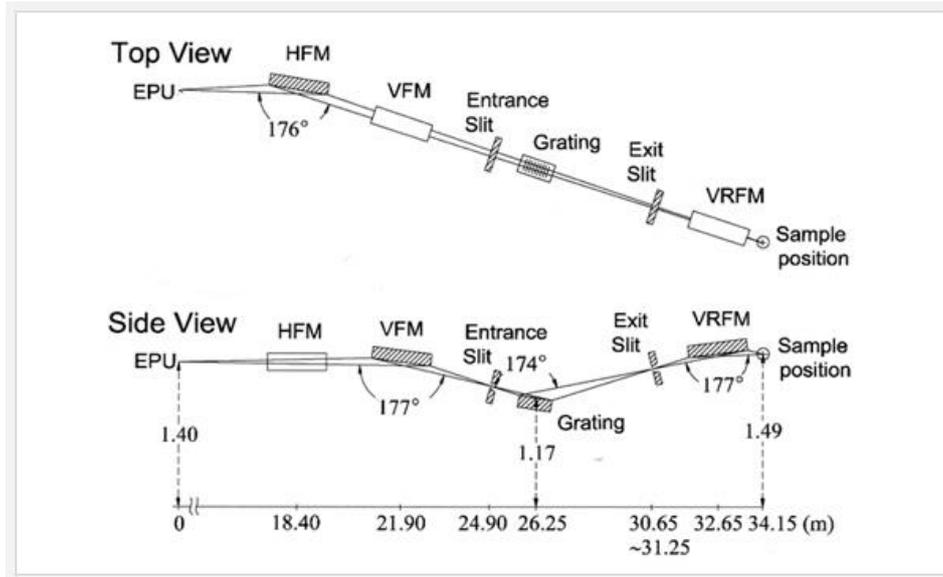

**Figure 8.** Magnetic spectroscopy beamline layout.

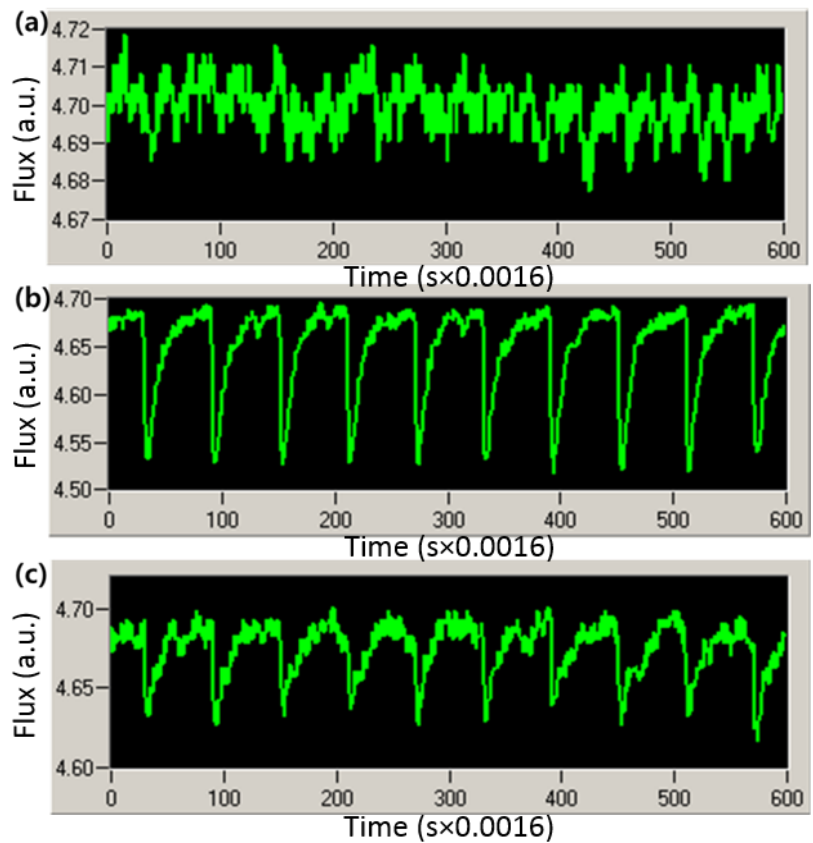

**Figure 9.** Measured mesh current at the beamline. Mesh current is acquired at 600 Hz. (a) No injection. (b) Main injection system on. (c) Main injection system and fast counter-kicker on.

**Table 1.** Specification of fast count kicker power supply.

| Parameter | Value | Unit |
| --- | --- | --- |
| Range of beam current shake (vertical) | 400 | $\mu m$ |
| Peak current | 50 | A |
| Supply voltage | 500 | V |
| Pulse width | 6.8 | $\mu s$ |
| Total inductance | 5.6 | µH |
| Total Capacitance | 2.4 | $\mu F$ |
| Total impedance | 10 | Ω |
| Energy per pulse | 0.3 | J |